\documentclass[10pt]{article}

\usepackage{amsfonts}
\usepackage{makeidx}
\usepackage{amssymb}
\usepackage{amsmath}
\usepackage{eurosym}
\usepackage{graphicx}

\begin{document}

\author{ {Mauro Fabrizio}\thanks{Dipartimento di Matematica, Universit\`a di Bologna, P.zza di Porta S.Donato 5, 40126 Bologna, Italy.
(e-mail address: {\tt fabrizio@dm.unibo.it})}}

\title{Superfluidity and vortices: A Ginzburg-Landau model \thanks{{Research performed
under the auspices of G.N.F.M. - I.N.D.A.M. and partially supported by  the Italian MIUR through the research project ``Mathematical models and methods in continuum physics'' and by University of Bologna through the project Mathematical Models of phase transitions for complex systems.}}}
\maketitle


\begin{abstract}
The paper deals with the study of superfluidity by a Ginzburg-Landau model
that investigates the material by a second order phase transition, in which
any particle has simultaneouly a normal and superfluid motion. This pattern
is able to describe the classical effects of superfluidity as the phase
diagram, the vortices, the second sound and the thermomechanical effect.
Finally, the vorticities and turbulence are described by an extension of the
model in which \ the material time derivative is used.
\end{abstract}

\noindent Keywords: 

Superconductivity, Ginzburg-Landau equations, critical fields, existence and uniqueness.

\section{ \ Introduction}

The behavior of superfluids is very different from the phenomenology of a
perfect fluid, as well as superconductivity \cite{G-L} is a phenomenon very
different from perfect conductivity (see Bardeen \cite{Bar}, Chandrasekhar 
\cite{Ch}, Landau \cite{landau}, \cite{Land}, London \cite{Lond1},
Mendelsshon \cite{Mend1}, Fabrizio,Gentili and Lazzari \cite{FGL}).
Meanwhile, there is an evident similarity between the behavior of
superfluids and superconductors. By the way, London claims in \cite{Lond1} 
\S\ 22 \textquotedblright ... in either case, superconductivity and
superfluid helium, the basic equations can be written in the same
form\textquotedblright , and Mendelsshon asserts "However, frictionless flow
and persistent currents are not the only features in which He II resembles a
superconductor. As at the transition of a metal to the superconductive
state, there occurs in helium also a rapid drop in the entropy. Moreover,
superfluid flow, just as a persistent supercurrent, is distinguished by zero
entropy". In view of the apparent analogies between superfluidity and
superconductivity, in this paper we present a phase transition model, which
use the two fluids Ginzburg-Landau view point, in which the equation
governing the motion of \ superfluid component $\mathbf{v}_{s}$ assumes an
analogous form as the equation for superconducting electrons. It turns out
in our analysis that $\nabla \times \mathbf{v}_{s}\neq 0$. Accordingly, the
equations allow to predict existence of vortices, which have been widely
studied by several authors (see: \cite{TT} cap. 6,\cite{Abri}, \cite{Ar-St}, 
\cite{Ar-St2}, \cite{Be-Ro}). Moreover in this paper, we characterize
superfluidity as a second order phase transition$\footnote{%
There is a marked difference between first-order and second-order phase
transitions. In first-order transitions, below the critical temperature the
whole body occurs in a single phase. Instead, in second-order transitions,
below the critical temperature the body consists of the simultaneous
occurrence of both phases. That is why we can view the material, below the
critical temperature, as consisting of both the normal fluid and the
superfluid (two fluid model).}$. Thus, our model for superfluidity is able
to explain the phase diagram represented in Fig.1, at least in the region
close to the $\lambda -$line, which is the most significant area. At the
same time, it allows to prove the existence of a critical velocity of
superflow, beyond which the frictionless flow breaks down.
\begin{center}
\bigskip \includegraphics[width=8cm]{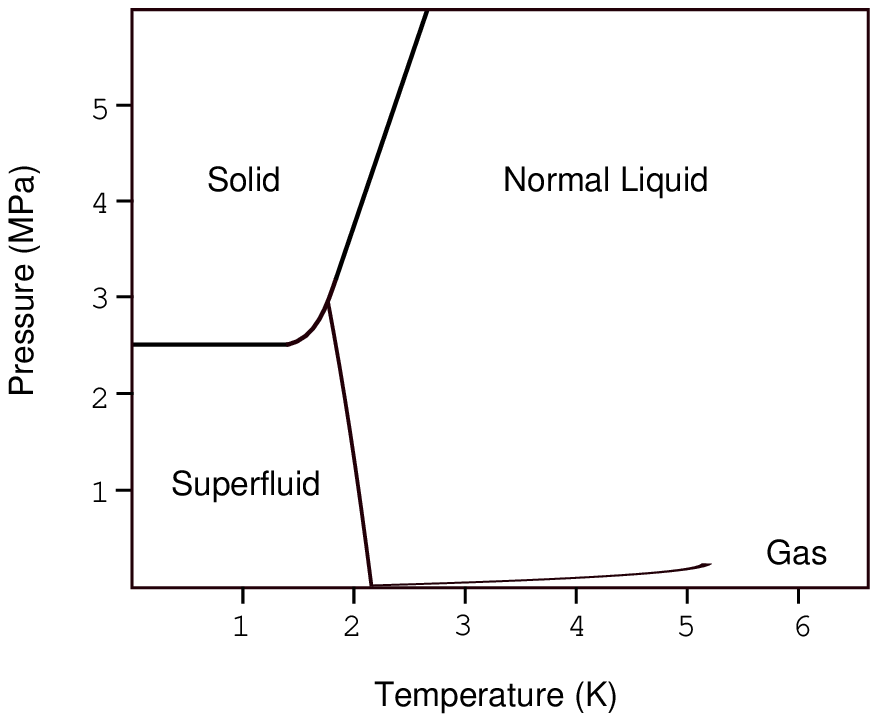}

Fig.1. Helium II. Phase diagram.
\end{center}
A remarkable feature of our model is the constraint on the rotational motion
by means of the equation $\nabla \times \mathbf{v}_{s}=\mathbf{v}_{n}$ which
relates $\mathbf{v}_{s}$ with the normal component of the velocity $\mathbf{v%
}_{n}.$ So that the velocity is defined by%
\begin{equation}
\mathbf{v}=\mathbf{v}_{n}+f^{2}\mathbf{v}_{s}.  \label{0}
\end{equation}%
The formula (\ref{0}) does not allow to define the superfluids by means of
two components, "the treatment of the liquid as a \textit{mixture} of normal
and superfluid \textit{parts} is simply a form of words ....Like any
description of quantum effects in classical terms, it is not entirely
adequate. It does not at all mean that the liquid can actually be separated
into parts" \cite{Lifs. Pit}. Actually as in (\ref{0}), any superfluid
particle is capable of two types of motion simultaneously, corresponding to
the energy spectrum of phonons and rotons. "One of these motions is \textit{%
normal}. i.e. has the same properties as that of an ordinary viscous liquid;
the other is \textit{superfluid}" \cite{Lifs. Pit}.

In Section 2 we will examine thermodynamical consistence of the model. Then,
in view of the similarity between superfluidity and superconductivity, we
will prove existence of vortices, whenever a superfluid, contained in a
cylindrical bucket, rotates rapidly around its axis. In particular, in
accordance with experimental results, we will show that, if the angular
velocity is low, the superfluid stays at rest. Instead, when the velocity
exceeds a threshold value, there occur vortices, similar to the vortices
observed in superconductivity. Moreover, the proposed model allows to
explain thermomechanical effect, in which we observe a motion of liquid
produced by heat, but in the same direction to heat flux. Finally, we prove
the existence of the phenomenon of second sound.

In section 3, we investigate the problem which arising when the flow is such
that we need to consider the material time derivative of the velocities.
This study introduces new non-linear terms that increase the complexity of
the model and require a suitable change of the previous pattern.

In the last part of the paper we study the connection between this model and
the turbulence. So that, as observed by Mendelsshon \cite{Mend1} "It is now
clear that the dependence of the heat conduction on the heat current
originally observed in Cambridge in 1937 ......is evidence of turbulence". A
remarkable feature of our model is the constraint on the rotational motion $%
\nabla \times \mathbf{v}_{s}=\mathbf{v}_{n},$ that relates $\mathbf{v}_{s}$
with the normal component of the velocity $\mathbf{v}_{n}$ and which
provides the behavior of a disordered set of thin vortices.

\section{A first model for superfluidity}

In this section we shall provide a first approximate model for the study of
superfluidity in a domain $\Omega \subset \mathbb{R}^{3}$ as a second order
phase transition, by means of the Ginzburg-Landau equation 
\begin{equation}
\frac{\partial f}{\partial t}=\frac{1}{\kappa }\nabla
^{2}f-f(f^{2}-1+u+\lambda p+\mathbf{v}_{s}^{2})  \label{1}
\end{equation}%
where $f$ denotes the order parameter (or phase field), while $p$ is the
pressure, $u$ is the absolute temperature and $\kappa ,\lambda $ are
positive constants. As studied in \cite{F1}, the Ginzburg-Landau equation
has to be considered as a new field equation, that we obtain by the balance
law on the structure order (See Appendix). Indeed, following Landau, we
suppose that the transition of the HeII from a normal to a superfluid state
induces a change in the internal structure order. The phase of this
transition is represented by the scalar parameter $f\in \left[ -1,1\right] $
that is linked with the density $n_{s}$ of the superfluid particles by the
formula 
\begin{equation*}
n_{s}=f^{2}.
\end{equation*}%
Hence the phase field $f^{2}=0$~denotes the normal state, while $f^{2}\in
\left( 0,1\right] $ describes a superfluid state. In this framework, in
equation (\ref{1}) if 
\begin{equation*}
R=u+\lambda p+\mathbf{v}_{s}^{2}-1>0,
\end{equation*}%
then the fluid is in the normal state, while when $R<0$ the pattern
describes the superfluid phase. In this work we do not employ the classical
two fluids model which introduces two different velocities and densities, as
for a mixture, but we suppose the velocity of any particle composed of two
excitements, the normal component $\mathbf{v}_{n}$ and the superfluid
component $\mathbf{v}_{s},$ such that 
\begin{equation}
\mathbf{v}=\mathbf{v}_{n}+f^{2}\mathbf{v}_{s}.  \label{2}
\end{equation}%
Because superfluidity as well as superconductivity must be studied as a
second order phase transition, then under the transition temperature, the
phase of some \ particles $f$ can be still equal to zero. Nevertheless,
according as the particles are in the normal phase ($f^{2}=0$) or in the
superfluid one ($f^{2}>0$ ), we cannot interpret this different behavior as
two different fluids.

Henceforth in this paper we choose $\lambda =0,$ and the density $\rho $ a
positive constant. In addition, in this first approximate model we suppose
the motion such that the acceleration $\mathbf{a}_{n}=\frac{\partial \mathbf{%
v}_{n}}{\partial t}$. Then, for the velocity $\mathbf{v}_{n}~$we propose the
modified Navier-Stokes equation 
\begin{equation}
\frac{\partial \mathbf{v}_{n}}{\partial t}=-\nabla p-\mu \nabla \times
\nabla \times \mathbf{v}_{n}-\mu \nabla \times f^{2}\mathbf{v}_{s}+\mathbf{g}
\label{3}
\end{equation}%
where $\mathbf{g}$ denotes the external force and $\mu $ is the viscosity
coefficient. Since we have supposed the superfluid as an incompressible
material, then the continuity equation provides 
\begin{equation}
\nabla \cdot \mathbf{v}_{n}=0  \label{4}
\end{equation}%
Furthermore, we suggest the following equations for the component $\mathbf{v}%
_{s}$ 
\begin{equation}
\frac{\partial \mathbf{v}_{s}}{\partial t}=-\nabla \phi -\mu \nabla \times
\nabla \times \mathbf{v}_{s}-\mu f^{2}\mathbf{v}_{s}+\nabla u+\mu \mathbf{h}
\label{5}
\end{equation}%
\begin{equation}
\nabla \cdot f^{2}\mathbf{v}_{s}=-\tau f^{2}\phi  \label{6}
\end{equation}%
where $\phi $ is a suitable scalar function and $\tau $ is a positive
constant. While the vector $\mathbf{h}$ is such that 
\begin{equation}
\mu \nabla \times \mathbf{h=g}-\nabla p~,~\nabla \cdot \mathbf{h=}0,\ \left. 
\mathbf{h\times n}\right\vert _{\partial \Omega }=0  \label{6a}
\end{equation}

Now, using (\ref{6a}), it is easy to prove that equation (\ref{3}) can be
obtained from equation (\ref{5}) if 
\begin{equation}
\nabla \times \mathbf{v}_{s}=\mathbf{v}_{n}  \label{2b}
\end{equation}%
Therefore, under the latter condition it will be sufficient to work only
with system (\ref{5})-(\ref{2b}), which contains equation (\ref{3}) as a
consequence.

\section{\protect\bigskip Thermodynamics}

In order to obtain the heat equation, let us consider the first law of
thermodynamics in the form 
\begin{equation}
\dot{E}=\mathcal{P}^{f}+\mathcal{P}^{\mathbf{v_{s}}}+h\,,  \label{3'}
\end{equation}%
where $E$ is the total energy and $h$ is the rate at which the heat is
absorbed by the material. The internal powers $\mathcal{P}^{f},\mathcal{P}^{%
\mathbf{v_{s}}},$ related to the variables $f,\mathbf{v}_{s},$ are given$%
\footnote{%
See Appendix.}$ by 
\begin{eqnarray}
\mathcal{P}^{f} &=&f_{t}^{2}+\frac{1}{2\kappa }\left[ (\nabla f)^{2}\right]
_{t}+\frac{1}{4}\left[ (1-f^{2})^{2}\right] _{t}+ff_{t}(u+\mathbf{v}_{s}^{2})
\label{4b} \\
\mathcal{P}^{\mathbf{v}_{s}} &=&\frac{1}{2}\left[ \mathbf{v}_{n}^{2}\right]
_{t}+\mu ^{-1}(\mathbf{v}_{st}+\nabla \phi -\nabla u)^{2}+f^{2}\mathbf{v}%
_{s}\cdot \mathbf{v}_{st}+u\nabla \cdot f^{2}\mathbf{v}_{s}+\tau f^{2}\phi
^{2}  \label{6d}
\end{eqnarray}%
From first law (\ref{3'}) we have 
\begin{equation}
h=c(u)u_{t}-f_{t}^{2}-\mu ^{-1}(\mathbf{v}_{st}+\nabla \phi - \nabla u%
)^{2}-uff_{t}-u\nabla \cdot f^{2}\mathbf{v}_{s}-\tau f^{2}\phi ^{2}\,,
\label{7}
\end{equation}%
where $c(u)$ is the specific heat, whose beaviour is represented in Fig.2.

\begin{center}
\includegraphics[width=8cm]{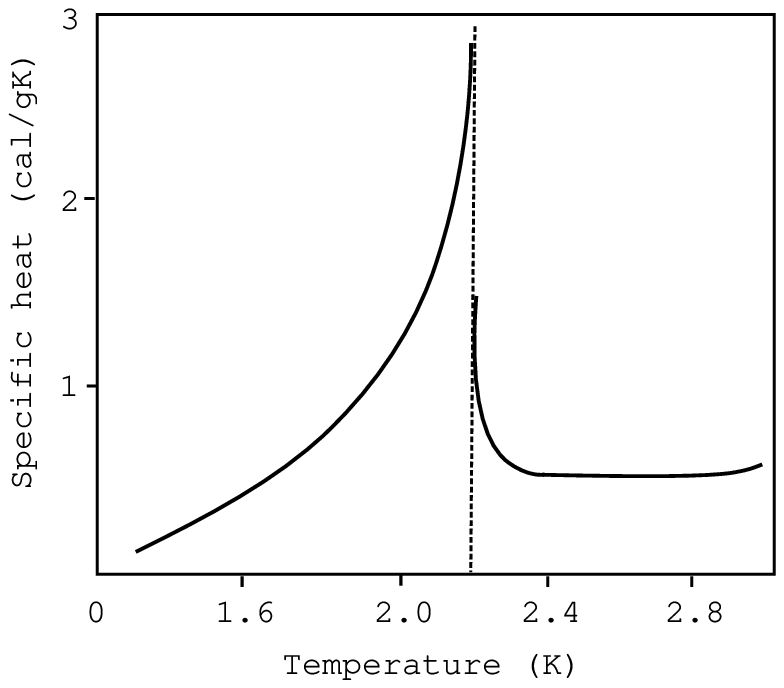}

Fig.2. Specific heat of Helium.
\end{center}

If $T=\frac{1}{2}\mathbf{v}_{n}^{2}+\frac{1}{2}f^{2}\mathbf{\ v}_{s}^{2}$
denotes the kinetic energy, and 
\begin{equation}
e=\int c(u)du+\frac{1}{4}(1-f^{2})^{2}+\frac{1}{2\kappa }(\nabla f)^{2}\,,
\label{7a}
\end{equation}%
is the internal energy, then the total energy $E=e+T$ \ is given by 
\begin{equation}
E=\frac{1}{2}\mathbf{v}_{n}^{2}+\frac{1}{2}f^{2}\mathbf{\ v}_{s}^{2}+\int
c(u)du+\frac{1}{4}(1-f^{2})^{2}+\frac{1}{2\kappa }(\nabla f)^{2}\,,
\label{8}
\end{equation}%
In our model the Fourier theory of heat conduction will not be modified.
Accordingly, the constitutive equation relating the heat flux $\mathbf{\ q}$
to the gradient of the temperature, assumes the classical form 
\begin{equation}
\mathbf{q}=-k(u)\nabla u\,,  \label{9}
\end{equation}%
where the conductivity $k$ depends on the absolute temperature. In this
framework, since the conductivity is very small when the absolute
temperature is close to zero, we suppose that 
\begin{equation}
k(u)=k_{0}u~,\text{~}~k_{0}>0\,.  \label{10}
\end{equation}%
The heat balance is expressed by the equation 
\begin{equation}
h=-\nabla \cdot \mathbf{\ q}+r\,,  \label{11}
\end{equation}%
from which we deduce the entropy equation (see \cite{Faf-Gio}) 
\begin{equation}
\frac{h}{u}-\frac{1}{u^{2}}\mathbf{\ q}\cdot \nabla u=-\nabla \cdot \frac{%
\mathbf{\ q}}{u}+\frac{r}{u}\,.  \label{12}
\end{equation}%
Therefore, 
\begin{equation}
K=\frac{h}{u}-\frac{1}{u^{2}}\mathbf{\ q}\cdot \nabla u  \label{13}
\end{equation}%
can be identified with the rate (per unit volume) at which entropy is
absorbed, while $\mathbf{\ q}/u$ and $r/u$ represent the entropy flux and
the entropy supply. Then equation (\ref{12}) describes the entropy balance 
\cite{Faf-Gio}, \cite{frem}. Because of the non-local constitutive model
given by system (\ref{1}-\ref{6}), we can state the following (see \cite%
{Faf-Gio}, \cite{FM}) \medskip \noindent

\textbf{Second law of thermodynamics }(non-local form)\textbf{.} \textit{%
\noindent There exists a function }$\eta (u,f)$\textit{, called entropy
function, such that} 
\begin{equation}
\frac{\partial \eta }{\partial t}\geq K+\nabla \cdot \mathbf{k}=\frac{h}{u}-%
\frac{1}{u^{2}}\mathbf{\ q}\cdot \nabla u\,+\nabla \cdot \mathbf{k}.
\label{14}
\end{equation}%
\textit{where }$\mathbf{k}$ is a suitable \textit{flux vector }such that $%
\left. \mathbf{k\cdot n}\right\vert _{\partial \Omega }=0$\textit{.}

\medskip It is easy to prove that system (\ref{1}-\ref{6}) with constitutive
equations (\ref{7}), (\ref{9}) agrees with the inequality (\ref{14}), if we
regard as entropy function the functional 
\begin{equation}
\eta (u,f)=\int \frac{c(u)}{u}du-G(f)\,.  \label{16}
\end{equation}%
where $G(f)=\frac{1}{2}f^{2}$and as flux vector $\mathbf{k=}f^{2}\mathbf{v}%
_{s}$. Finally, in the sequel we shall consider an approximation of (\ref{12}%
), namely 
\begin{equation}
\frac{c(u)}{u}u_{t}-\dot{G}(f)=\nabla \cdot (f^{2}\mathbf{\ v}%
_{s})+k_{0}\nabla ^{2}u+\frac{r}{u}\,,  \label{17}
\end{equation}%
where we have neglected the dissipative terms $f_{t}^{2},(\mathbf{v}%
_{st}+\nabla \phi -\nabla u)^{2},(\nabla u)^{2},\tau f^{2}\phi ^{2}$.

Now, let us evaluate the free energy $\psi $ defined by 
\begin{equation*}
\psi =e-u\eta \,.
\end{equation*}%
From expressions (\ref{7a}) and (\ref{16}) of the internal energy and the
entropy, it follows that 
\begin{equation}
\psi =\frac{1}{2\kappa }(\nabla f)^{2}+\frac{1}{4}(1-f^{2})^{2}+uG(f)+(1-u)%
\int \frac{c(u)}{u}du\,.  \label{17bis}
\end{equation}%
We will prove that system (\ref{1}), (\ref{2}), (\ref{3}), (\ref{17}) allows
to describe the main aspects of superfluidity. First, we examine the phase
diagram represented in fig.1. If we focus our attention on phenomena in
which the pressure is below 25 atm, the transition occurs along the $\lambda
-$line, which we can assume close to a vertical line. This phenomenon is
explained by equation (\ref{1}). Indeed, by determining the transition
states $(u_{T},\mathbf{v}_{sT})$, we observe that the line which separates
the normal and the superfluid phases is given by the curve 
\begin{equation}
u+\lambda p+\mathbf{v}_{s}^{2}=1  \label{18}
\end{equation}%
If we suppose $\lambda \ll 1$ and consider the equilibrium states, i.e. $%
\mathbf{v}_{s}=0,$ then from (\ref{18}) we get the equation $u=1,$ which is
a good approximation of the $\lambda -$line (see fig.1). Instead, when $%
\mathbf{v}_{s}\neq 0,$ the model is naturally able to account for the
existence of a critical velocity, namely a threshold value above which the
frictionless flow breaks down. This follow directly from equation (\ref{18}%
). \bigskip\ 

The latent heat $l$~related to the transition is given by 
\begin{equation*}
l=u_{T}(\eta (u_{T},f_{1})-\eta (u_{T},f_{2}))
\end{equation*}%
where $f_{1~}$and $f_{2}$ are the local minima of the free energy at the
transition points $u+\mathbf{v}_{s}^{2}=1$, In this case $f_{1}=f_{2}=0$.
Hence, $l=0$, which means that the transition is of second order.\bigskip\ 

In summary, our differential system assumes the form 
\begin{equation}
\frac{\partial f}{\partial t}=\frac{1}{\kappa }\nabla ^{2}f-f(f^{2}-1+u+%
\mathbf{v}_{s}^{2})  \label{181}
\end{equation}%
\begin{equation}
\frac{\partial \mathbf{v}_{s}}{\partial t}=-\nabla \phi -\mu \nabla \times
\nabla \times \mathbf{v}_{s}-\mu f^{2}\mathbf{v}_{s}+\nabla u+\mu \mathbf{h}
\label{184}
\end{equation}%
\begin{equation}
\nabla \cdot f^{2}\mathbf{v}_{s}=-\tau f^{2}\phi  \label{185}
\end{equation}%
\begin{equation}
\frac{c(u)}{u}\frac{\partial u}{\partial t}-f\frac{\partial f}{\partial t}%
=\nabla \cdot (f^{2}\mathbf{\ v}_{s})+k_{0}\nabla ^{2}u+\frac{r}{u}
\label{186}
\end{equation}%
Moreover, we associate to system (\ref{181}-\ref{186}) the boundary
conditions on the domain $\Omega $ 
\begin{eqnarray}
\nabla f\cdot \mathbf{n}|_{\partial \Omega } &=&0\,,~~\ (\nabla \times 
\mathbf{\ v}_{s})\times \mathbf{\ n}|_{\partial \Omega }=\mathbf{\ \omega }
\label{187} \\
\left. \mathbf{v}_{s}\cdot \mathbf{\ n}\right\vert _{\partial \Omega }
&=&0,~\left. \mathbf{v}_{n}\right\vert _{\partial \Omega }=0,~\left.
u\right\vert _{\partial \Omega }=u_{b}  \label{188}
\end{eqnarray}%
It is worth noting that, when $f=0$, in view of (\ref{2}), $\mathbf{\ v}=%
\mathbf{\ v}_{n}$ and equation (\ref{5}) is inessential, since the motion is
described by means of the Navier-Stokes equations 
\begin{equation}
\frac{\partial \mathbf{v}}{\partial t}=-\nabla p-\mu \nabla \times \nabla
\times \mathbf{v}+\mathbf{g~},~\text{~}~\text{~}\nabla \cdot \mathbf{\ v}=0,
\label{19}
\end{equation}%
obtained from (\ref{3}),(\ref{4}). On the contrary, when the transition
occurs, $f>0$ and the consequent breaking of symmetry leads to system (\ref%
{181})-(\ref{186}). Well-posedness of differential problem (\ref{181})-(\ref%
{186}) in the case $f=0$ has to be investigated carefully in order to prove
uniqueness of solutions. Indeed, when $f=0$ equation (\ref{184}) reads 
\begin{equation}
\mathbf{\ }\frac{\partial \mathbf{v}_{s}}{\partial t}=-\nabla \phi -\mu
\nabla \times \nabla \times \mathbf{v}_{s}-\nabla u+\mu \mathbf{h},
\label{*}
\end{equation}%
while (\ref{185}) is satisfied identically. The temperature $u$ is
univocally determined through (\ref{186}), endowed with suitable initial and
boundary conditions. However, since equation (\ref{*}) is no longer coupled
with (\ref{185}), uniqueness of solutions cannot be proved. For this reason,
it is necessary a choice of a gauge, which consists in decomposing the
velocity $\mathbf{\ v}_{s}$ as 
\begin{equation*}
\mathbf{\ v}_{s}=\mathbf{\ A}+\nabla \varphi ,~or\ \nabla \times \mathbf{A}=%
\mathbf{v}_{n}
\end{equation*}%
The vector$\mathbf{\ A}$ is specified by fixing a value for its divergence
and choosing suitable boundary conditions. By analogy with
superconductivity, we will choose London's gauge 
\begin{equation*}
\nabla \cdot \mathbf{\ A}=0\,,~\ \ \ \ \ \mathbf{\ A}\cdot \mathbf{\ n}%
|_{\partial \Omega }=0\,.
\end{equation*}%
Therefore, from (\ref{*}) we deduce%
\begin{equation}
\frac{\partial \mathbf{A}}{\partial t}\mathbf{=}-\nabla (\phi +\dot{\varphi}%
)-\mu \nabla \times \nabla \times \mathbf{A}-\nabla u+\mu \mathbf{h}
\label{21}
\end{equation}%
\bigskip Finally, from equation (\ref{21}) we get equation (\ref{19}). The
main result which this model allows to prove is the presence of quantized
vortices, when a cylindrical bucket containing Helium II rotates around its
axis. Consider the stationary problem, in addition let us suppose the
temperature gradient $\nabla \tilde{u}=0$. Hence the differential system (%
\ref{181})-(\ref{185}) with $\mathbf{h}=0$ assumes the form 
\begin{eqnarray}
\frac{1}{\kappa }\nabla ^{2}f-f(f^{2}-1+\tilde{u}+\gamma \mathbf{v}_{s}^{2})
&=&0  \label{24} \\
-\nabla \phi -\mu \nabla \times \nabla \times \mathbf{v}_{s}-\mu f^{2}%
\mathbf{v}_{s} &=&0  \label{25} \\
\nabla \cdot f^{2}\mathbf{\ v}_{s}+\tau f^{2}\phi &=&0  \label{27}
\end{eqnarray}%
together with the boundary conditions 
\begin{equation}
\nabla f\cdot \mathbf{n}|_{\partial \Omega }=0\,,~~(\nabla \times \mathbf{v}%
_{s})\times \mathbf{\ n}|_{\partial \Omega }=\mathbf{\ \omega ~,}\,~~\mathbf{%
\ v}_{s}\cdot \mathbf{\ n}|_{\partial \Omega }=0.  \label{27b}
\end{equation}%
If the angular velocity is below a critical value $\mathbf{\omega }_{c_{1}}$%
, depending on the geometry of the problem, the superfluid component stays
at rest, with the exception of a very small penetration depth close to the
boundary$\footnote{%
Consider a fluid in a long straight wire of circular cross-section carrying
a rotation around the axis. We suppose cylindrical symmetry and indroduce
the coordinates $z,r,\theta ~(\mathbf{k,r,j)}$. Then we have 
\begin{eqnarray*}
\mathbf{v}_{s}(r) &=&v_{\theta }(r)\mathbf{j} \\
\phi &=&\phi (r) \\
\mathbf{v}_{n}(r) &=&\nabla \times \mathbf{v}_{s}(r)=\frac{\partial
v_{\theta }(r)}{\partial r}\mathbf{k}
\end{eqnarray*}%
Therefore, from (\ref{25}) we obtain%
\begin{eqnarray*}
\nabla \phi (r) &=&0 \\
\frac{d^{2}v_{\theta }}{dr^{2}} &=&-\frac{1}{r}\frac{dv_{\theta }}{dr}%
+(f^{2}+1)v_{\theta }
\end{eqnarray*}%
The solutions related with a given field $v_{0}$ at the surface of the
cylinder of radius $R$ is%
\begin{equation*}
v_{\theta }=v_{0}\frac{I_{1}(r)}{I_{1}(R)}
\end{equation*}%
where $I_{1}$ is the modified Bessel functions of imaginary argument.
Because of the behavior of $I_{1}(r)$ for $r<R$, the velocity $v_{\theta
}(r) $ will be confined to a thin layer next to the boundary.}$. This
equilibrium state is called Landau's state. The phenomenon is analogous to
the diamagnetism shown in superconductivity by the Meissner effect, when for
a second type superconductor the magnetic field is lower that a given field $%
H_{c_{1}}$. When the angular velocity is greater that $\mathbf{\omega }%
_{c_{1}}$, we observe the creation of vortex lines in an exactly similar way
as in superconductivity. Since, problem (\ref{24})-(\ref{27b}) is identical
to the Ginzburg-Landau system in stationary case, studied by Abrikosov \cite%
{Abri}, for superconductors of type II. Therefore, when $\mathbf{\omega }>%
\mathbf{\omega }_{c1}$ with the same analysis, we may prove the existence of
vortex lines in a superfluid. Moreover, as observed in \cite{TT}, when the
rotation velocity exceeds a second threshold value $\mathbf{\omega }_{c2}$,
owing to increase of vortices and to the overlapping of their cores,
superfluidity is destroyed. Even this effect is analogous with the
corresponding phenomena in superconductivity and it can be explained by (\ref%
{21}). Indeed, for sufficiently large values of $\mathbf{\ v}_{s}$ we have 
\begin{equation*}
\tilde{u}+\mathbf{\ v}_{s}^{2}>1\,,
\end{equation*}%
so that the fluid is in the normal phase. In every practical situations,
angular velocities $\mathbf{\omega }$ satisfy the condition 
\begin{equation*}
\mathbf{\omega }<\mathbf{\omega }_{c2}\,.
\end{equation*}%
Moreover, the first critical velocity is such that 
\begin{equation*}
\mathbf{\omega }_{c1}\ll 1\,.
\end{equation*}%
Another important phenomenon occurring in superfluidity, it is the
thermomechanical effect according to which the particles of the superfluid
flow in the same direction to heat flux. This phenomenon, which contradicts
the classical behavior of thermo-fluid mechanics, can be explained by means
of equation (\ref{5}). Indeed in (\ref{5}) the sign of $\nabla u$ is the
same as the acceleration $\mathbf{v}_{st}$ of the superfluid component. This
means that the gradient of the temperature causes an increase of the
velocity $\mathbf{\ v}_{s}$ in the same direction. Moreover, to understand
better the phenomenon, we have to observe that in the thermo-mechanical
effect the tube may be straight and very narrow. As a consequence, the
motion of the superfluid phase $(f\neq 0)$ through the tube is required to
satisfy $\nabla \times $ $\mathbf{v}_{s}=\mathbf{v}_{n}=0$ and hence the
velocity is given by $\mathbf{v}=f^{2}\mathbf{v}_{s}$. In other words, since
the normal component is absent, the motion is necessarily that of the
superfluid component. This in turn allows for the flow through very narrow
tubes. It follows from (\ref{5}) that, in stationary conditions, the motion
is governed by the equation 
\begin{equation}
-\nabla \phi -f^{2}\mathbf{v}_{s}+\nabla u+\mathbf{h}=0,  \label{30a}
\end{equation}%
which, consistent by (\ref{30a}), shows that the term $\nabla u~$ favors the
particle displacement toward regions at a higher temperature. Moreover,
because of the narrowness of the tube, only superfluid particles ($f=0$) are
allowed to flow in. Instead the normal component undergoes to a viscous
resistance which forbids the flow to cross the tube.

Finally, we study the second sound effect. For this problem, let us suppose $%
f=const.$ and the supplies $\mathbf{h}=0,~r=0$. Then, from (\ref{184})-(\ref%
{186}) it follows that 
\begin{equation}
\nabla \cdot \mathbf{v}_{st}=-\nabla ^{2}\phi -\mu \nabla \cdot f^{2}\mathbf{%
v}_{s}+\nabla ^{2}u  \label{30b}
\end{equation}%
\begin{equation}
\alpha u_{tt}=\nabla \cdot (f^{2}\mathbf{\ v}_{st})+k_{0}\nabla ^{2}u_{t}
\label{30c}
\end{equation}%
where $\alpha =\frac{c(u)}{u}$ will be supposed a positive constant.
Moreover, in the superfluid framework we can suppose the thermal
conductivity $k_{0}\ll 1,$therefore we get from (\ref{27}) and (\ref{30c}) 
\begin{equation*}
\alpha u_{tt}-f^{2}\nabla ^{2}u=f^{2}(\tau \mu f^{2}\phi -\nabla ^{2}\phi )
\end{equation*}%
In other words, the temperature satisfies a wave equation that is able to
provide a new kind of wave.

\section{General case}

In the previous sections the material time derivative $\frac{dF}{dt}=\frac{%
\partial F}{\partial t}+\mathbf{v\cdot \nabla }F~$was always completely
ignored, because we have supposed slow motions. In this section we shall
study the general case and then we use the material time derivative $\frac{dF%
}{dt}$ which we denote only with $\dot{F}$. In addition we suppose the fluid
incompressible with a given constant density $\rho =1$.\ Then, on the domain 
$\Omega ,~$the differential system assumes the new form 
\begin{equation}
\dot{f}=\frac{1}{\kappa }\nabla ^{2}f-f(f^{2}-1+u+\mathbf{v}_{s}^{2})
\label{31}
\end{equation}%
\begin{equation}
\mathbf{\dot{v}}_{n}=-\nabla p-\mu \nabla \times \nabla \times \mathbf{v}%
_{n}-\mu \nabla \times f^{2}\mathbf{v}_{s}+\mathbf{g}  \label{33}
\end{equation}%
\begin{equation*}
\nabla \cdot \mathbf{v}_{n}=0
\end{equation*}%
\begin{equation}
\mathbf{\dot{v}}_{s}=-\nabla \phi -\mu \nabla \times \mathbf{v}_{n}-\mu f^{2}%
\mathbf{v}_{s}+\nabla u+\mu \mathbf{h}  \label{34a}
\end{equation}%
\begin{equation}
\nabla \cdot f^{2}\mathbf{v}_{s}=-\tau f^{2}\phi  \label{35}
\end{equation}

Because of the use of the material time derivative, even if we suppose $\mu
\nabla \times \mathbf{h=g}-\nabla p$, it is not possible to prove that from
equations (\ref{33}) and (\ref{34a}) we obtain equation (\ref{3}).
Otherwise, if we consider the {curl} of the equation (\ref{34a}) and
compare this new equation with (\ref{33}), we have the restriction 
\begin{equation}
\nabla \times \mathbf{\dot{v}}_{s}=\mathbf{\dot{v}}_{n}  \label{39e}
\end{equation}%
\ Moreover, the internal powers $\mathcal{P}^{f},\mathcal{\ \ P}^{\mathbf{%
v_{s}}},$ assume the analogous form%
\begin{eqnarray}
\mathcal{P}^{f} &=&\dot{f}^{2}+\frac{1}{2\kappa }\left[ (\nabla f)^{2}\right]
^{\cdot }+\frac{1}{4}\left[ (1-f^{2})^{2}\right] ^{\cdot }+f\dot{f}(u+%
\mathbf{v}_{s}^{2})  \label{37} \\
\mathcal{P}^{\mathbf{v}_{s}} &=&\mu \nabla \times \mathbf{v}_{n}\cdot 
\mathbf{\dot{v}}_{s}+\mu ^{-1}(\mathbf{\dot{v}}_{s}+\nabla \phi -\nabla
u)^{2}+f^{2}\mathbf{v}_{s}\cdot \mathbf{\dot{v}}_{s}+u\nabla \cdot f^{2}%
\mathbf{v}_{s}+\tau f^{2}\phi ^{2}  \label{39}
\end{eqnarray}%
In particular, from (\ref{33}-\ref{34a}) we have

\begin{equation}
\begin{array}{c}
0=\int_{\Omega }(\frac{1}{2}\left[ \mathbf{v}_{n}^{2}\right] ^{\cdot }+\mu
(\nabla \times \mathbf{v}_{n}+f^{2}\mathbf{v}_{s})\cdot \nabla \times 
\mathbf{v}_{n}-(\mathbf{g}-\nabla p)\cdot \mathbf{v}_{n})dx \\ 
=\int_{\Omega }(\frac{1}{2}\left[ \mathbf{v}_{n}^{2}\right] ^{\cdot }-\nabla
\times \mathbf{v}_{n}\cdot \mathbf{\dot{v}}_{s})dx%
\end{array}
\label{39c}
\end{equation}%
Hence, if (\ref{2b}) holds the identity (\ref{39c}) is satisfied, but the
vice versa is not true. However, from (\ref{39c}) and (\ref{188}) we obtain
the restriction%
\begin{equation}
\int_{\Omega }(\frac{1}{2}\left[ \mathbf{v}_{n}^{2}\right] ^{\cdot }-\nabla
\times \mathbf{\dot{v}}_{s}\cdot \mathbf{v}_{n})dx=0  \label{39d}
\end{equation}%
which is satisfied by means of (\ref{39e}).

\bigskip

Then, from the first law (\ref{3'}), we get 
\begin{eqnarray}
\dot{e} &=&h+\dot{f}^{2}+\frac{1}{2\kappa }\left[ (\nabla f)^{2}\right]
^{\cdot }+\frac{1}{4}\left[ (1-f^{2})^{2}\right] ^{\cdot }+f\dot{f}(u+%
\mathbf{v}_{s}^{2})+\nabla \times \mathbf{v}_{n}\cdot \mathbf{\dot{v}}_{s}+
\label{41a} \\
&&\mu ^{-1}(\mathbf{\dot{v}}_{s}+\nabla \phi -\nabla u)^{2}+f^{2}\mathbf{v}%
_{s}\cdot \mathbf{\dot{v}}_{s}+u\nabla \cdot f^{2}\mathbf{v}_{s}+\tau
f^{2}\phi ^{2}  \notag
\end{eqnarray}%
where the internal energy is given by 
\begin{equation}
e=\int c(u)du+\frac{1}{4}(1-f^{2})^{2}+\frac{1}{2\kappa }(\nabla f)^{2}\,,
\label{41}
\end{equation}
which implies 
\begin{equation}
h=c(u)\dot{u}-\dot{f}^{2}-(\mathbf{\dot{v}}_{s}+\nabla \phi -\nabla u)^{2}-uf%
\dot{f}-u\nabla \cdot f^{2}\mathbf{v}_{s}-\tau f^{2}\phi ^{2}\,,  \label{40}
\end{equation}%
where $c(u)=e_{u}(u)$ is the specific heat.

With similar arguments considered in the previous section we obtain the same
function for the entropy 
\begin{equation}
\eta (u,f)=\int \frac{c(u)}{u}du-G(f)\,.  \label{16a}
\end{equation}%
and under the same approximations, the following heat equation 
\begin{equation}
\frac{c(u)}{u}\dot{u}-\dot{G}(f)=\nabla \cdot (f^{2}\mathbf{\ v}%
_{s})+k_{0}\nabla ^{2}u+\frac{r}{u}\,,  \label{17a}
\end{equation}%
From expressions (\ref{8}) and (\ref{16a}) of the internal energy and the
entropy, it follows that 
\begin{equation}
\psi =\frac{1}{2\kappa }(\nabla f)^{2}+\frac{1}{4}(1-f^{2})^{2}+uG(f)+(1-u)%
\int \frac{c(u)}{u}du\,.  \label{17bisa}
\end{equation}%
We will prove that system (\ref{31})-(\ref{35}), (\ref{17a}) allows to
describe the main aspects of superfluidity, but in more complex form,
because of the presence of total time derivative and of equation (\ref{39d}).

\section{Turbulence}

It is well known that turbulence is a phenomenon which has been observed
quite frequently in superfluidity. "Hydrodynamic flow in both classical and
quantum fluids can be either laminar or turbulent. To describe the latter,
vortices in turbulent flow are modelled with stable vortex filaments. While
this is an idealization in classical fluids, vortices are real topologically
stable quantized objects in superfluids. Thus superfluid turbulence is
thought to hold the key to new understanding on turbulence in general." \cite%
{Be-Ro}, (see also \cite{FABE}).

This effect is connected with the definition (\ref{2}) of velocity and with
the relation (\ref{2b}) or (\ref{39d}) between normal and superfluid
components. In particular, if we consider the model described by the system (%
\ref{1}), (\ref{2}), (\ref{3}), (\ref{17}), then we observe that, under the
critical temperature $T_{c}$, the path of the particles assumes an
helicoidal motion, \ whose circulation, defined around the vortex core, is
quantized. If the temperature goes below a second critical temperature $%
T_{c2},$ the number of the superfluid particles increases. In such a case
the approximation (\ref{1}), (\ref{2}), (\ref{3}), (\ref{17}) is not
correct, because we need to consider the time material derivative. Hence the
motion assume a chaotic form, typical of the turbulence flow. Otherwise, it
is possible to approximate the system (\ref{31}-\ref{35}), (\ref{17a})
considering the material time derivative for $\mathbf{v}_{n}$ and $\mathbf{v}%
_{s}$, then we obtain the system

\begin{equation}
\frac{\partial f}{\partial t}=\frac{1}{\kappa }\nabla ^{2}f-f(f^{2}-1+u+%
\mathbf{v}_{s}^{2})  \label{28.1}
\end{equation}%
\begin{equation}
\mathbf{\dot{v}}_{n}=-\nabla p-\mu \nabla \times \nabla \times \mathbf{v}%
_{n}-\mu \nabla \times f^{2}\mathbf{v}_{s}+\mathbf{g}  \label{28.2}
\end{equation}%
\begin{equation}
\nabla \times \mathbf{\dot{v}}_{s}=\mathbf{\dot{v}}_{n}~,~\nabla \cdot 
\mathbf{v}_{n}=0  \label{28.3}
\end{equation}%
\begin{equation}
\mathbf{\dot{v}}_{s}=-\nabla \phi -\mu \nabla \times \mathbf{v}_{n}-\mu f^{2}%
\mathbf{v}_{s}+\nabla u+\mu \mathbf{h}  \label{28.4}
\end{equation}%
\begin{equation}
\nabla \cdot f^{2}\mathbf{v}_{s}=-\tau f^{2}\phi  \label{2845}
\end{equation}%
\begin{equation}
c_{0}\frac{\partial u}{\partial t}-\dot{G}(f)=\nabla \cdot (f^{2}\mathbf{\ v}%
_{s})+k_{0}\nabla ^{2}u+\frac{r}{u}  \label{28.5}
\end{equation}%
where in this framework we have supposed $\frac{c(u)}{u}=c_{0}$ positive
constant. This system is completed by the boundary conditions (\ref{187}-\ref%
{188}).

Now, let us consider the phenomenon of turbulence in viscous fluids. As for
a superfluid we suppose the velocity given by (\ref{2}). \ While, we suggest
for the differential problem a first order phase transition model by the
following system

\begin{equation}
\dot{f}=\frac{1}{\kappa }\nabla ^{2}f-\frac{1}{2}(N_{R}^{2}F^{\prime }(f)+%
\mathbf{v}_{s}^{2}G^{\prime }(f))  \label{29.1}
\end{equation}%
\begin{equation}
\dot{\mathbf{v}}_{n}=-\nabla p-\mu \nabla \times \nabla \times \mathbf{v}%
_{n}-\mu \nabla \times G(f)\mathbf{v}_{s}+\mathbf{g}  \label{29.3}
\end{equation}%
\begin{equation}
\nabla \times \mathbf{\dot{v}}_{s}=\mathbf{\dot{v}}_{n}~,~~\nabla \cdot 
\mathbf{v}_{n}=0  \label{29.2}
\end{equation}%
\begin{equation}
\dot{\mathbf{v}}_{s}=-\nabla \phi -\mu \nabla \times \mathbf{v}_{n}-\mu G(f)%
\mathbf{v}_{s}+\mu \mathbf{h}  \label{29.4}
\end{equation}%
\begin{equation}
\nabla \cdot G(f)\mathbf{v}_{s}=-\nu G(f)\phi  \label{29.5}
\end{equation}%
where $\mathbf{g}$ and $\mathbf{h}$ are related by (\ref{6a}), $N_{R}$ is a
constant related with the Reynolds number and $\nu $ a positive constant.
Moreover, because of the first order phase transition described by the
system (\ref{29.1}-\ref{29.5}), the functions $F(f)$ and $G(f)$ are now
given by the following fourth order polynomials 
\begin{equation}
F(f)=\frac{f^{4}}{4}-\frac{f^{3}}{3}~,~\text{~}G(f)=\frac{f^{4}}{4}-\frac{%
2f^{3}}{3}+\frac{f^{2}}{2}  \label{30}
\end{equation}

Finally, in this framework the internal powers connected with the variables $%
f,$ and $\mathbf{v}_{s}$ are given by 
\begin{eqnarray}
\mathcal{P}^{f} &=&\dot{f}^{2}+\frac{1}{2\kappa }\left[ (\nabla f)^{2}\right]
^{\cdot }+N_{R}\dot{F}(f)+\dot{G}(f)\mathbf{v}_{s}^{2}  \label{301} \\
\mathcal{P}^{\mathbf{v}_{s}} &=&G(f)\mathbf{v}_{s}\cdot \mathbf{\dot{v}}%
_{s}+\mu ^{-1}(\mathbf{\dot{v}}_{s}+\nabla \phi )^{2}+\frac{1}{2}\left[ 
\mathbf{v}_{n}^{2}\right] ^{\cdot }  \label{302} \\
\mathcal{P}^{\mathbf{v}_{n}} &=&\mu (\nabla \times \mathbf{v}_{n})^{2}+G(f)%
\mathbf{v}_{s}\cdot \mathbf{\nabla \times v}_{n}+\frac{1}{2}\left[ \mathbf{v}%
_{n}^{2}\right] ^{\cdot }  \label{303}
\end{eqnarray}

For this isothermal problem the laws of thermodynamics are given by the
following

\bigskip \textbf{Dissipation Principle. }There exists a functional $\psi (f,%
\mathbf{v}_{s})$, called free energy, such that%
\begin{equation*}
\dot{\psi}\leq \dot{f}^{2}+\frac{1}{2\kappa }\left[ (\nabla f)^{2}\right]
^{\cdot }+N_{R}\dot{F}(f)+\mu ^{-1}(\mathbf{\dot{v}}_{s}\mathbf{+}\nabla
\phi )^{2}
\end{equation*}

Then, the free energy is function only of $f$ and is defined by%
\begin{equation*}
\psi (f)=\frac{1}{2\kappa }(\nabla f)^{2}+N_{R}F(f)
\end{equation*}

\section{Appendix}

In many natural phenomena, as phase transitions, chemical reactions,
biological (processes) models, we can observe a natural evolution toward
change of material structure order. We suggest that this behavior is related
with a natural process which we may represent as a balance law\ on the
structure order and it will be given as a function of order parameter $f$ .
Moreover, as for the other field equations, to this new law we may connect
an internal power, which must be considered in the energy balance law. In
this framework the structure order is a new form of energy, because during
the transformation we observe a variation of the structural energy. \
Actually, if we consider a phase transition we notice a transformation from
a less ordered material structure to a more ordered one or vice versa (see 
\cite{F1}, \cite{Fr-Gur}). Moreover, below a critical temperature, the
structure order of many materials is greater then above. We meet, analogous
behaviors, during biological precesses and chemical reactions, always
connected with structure order variations.

In order to obtain a balance law on the structure order. Consider a body $%
\mathcal{B}$, for any sub-body $S\subset \mathcal{B}$, we denote with $%
\mathcal{S}^{i}(S)\mathcal{\ }$the rate at which structure order is absorbed
by the material per unit time, given by 
\begin{equation}
\mathcal{S}^{i}(S)=\int_{S}\rho kdv
\end{equation}%
where $\rho $ is the density and $k$ the \textit{internal specific structure
order}. While, the \textit{external order structure} $\mathcal{S}^{e}(S)$
assumes the form 
\begin{equation}
\mathcal{S}^{e}(S)=\int_{\partial S}\mathbf{p}\cdot \mathbf{n}%
ds+\int_{S}\rho \delta dv
\end{equation}%
where the vector $\mathbf{p}$ denotes the \textit{order structure flux} and $%
\delta $ the structure order supply.

Hence, the order structure balance is given for any $S\subset \mathcal{B}$
by the equality 
\begin{equation}
\int_{S}\rho kdv=\int_{\partial S}\mathbf{p}\cdot \mathbf{n}ds+\int_{S}\rho
\delta dv  \label{1.3}
\end{equation}%
In local form the integral equality (\ref{1.3}) implies the identity 
\begin{equation}
\rho k=\nabla \cdot \mathbf{p}+\rho \delta  \label{1.4}
\end{equation}%
In any model of phase transitions the functions $k$ and $\mathbf{p}$ are
usually defined by 
\begin{equation}
k=f_{t}+F^{\prime }(f)+mG^{\prime }(f)  \label{1.4b}
\end{equation}%
\begin{equation}
\mathbf{p}=\frac{1}{\kappa }\nabla f  \label{1.4c}
\end{equation}%
where $\ m$ is a suitable coefficient depending of the field. While $F$ and $%
G$ are functions that characterize the order and the feature of the
transition. For a second order phase transition as superfluidity $F$ and $G$
are defined by 
\begin{equation}
F(f)=\frac{f^{4}}{4}-\frac{f^{2}}{2}~,~\text{~}G(f)=\frac{f^{2}}{2}
\label{1.4d}
\end{equation}%
while $m=u+\lambda p+\mathbf{v}_{s}^{2},$ from which we obtain by (\ref{1.4}%
), (\ref{1.4b}) and (\ref{1.4c}) the equation%
\begin{equation}
\rho f_{t}=\frac{1}{\kappa }\nabla ^{2}f-\rho F^{\prime }(f)-\rho (u+\lambda
p+\mathbf{v}_{s}^{2})G^{\prime }(f)  \label{1.4a}
\end{equation}%
i.e. the equation (\ref{1}) with $\rho =1$

Because the equation (\ref{1.4}) is joined with an energy, we have to
introduce the \textit{power balance} connected with this equation, namely 
\begin{equation}
\rho kf_{t}+\mathbf{p}\cdot \nabla f_{t}=\nabla \cdot (f_{t}\mathbf{p)+}\rho
f_{t}\delta  \label{1.5}
\end{equation}

Finally, we denote by 
\begin{equation}
\mathcal{P}^{f}=\rho kf_{t}+\mathbf{p}\cdot \nabla f_{t}\;,\;\;\mathcal{\ P}%
_{e}^{f}=\nabla \cdot (f_{t}\mathbf{p)+}\rho f_{t}\delta   \label{1.6}
\end{equation}%
the \textit{internal} and the\textit{\ external order structure power} 
\textit{density }respectively.

\end{document}